# The AI Criminal Mastermind

Joshua Krook

Era AI Fellow / University of Antwerp

## 1. Introduction

In this paper, I evaluate the risks of an AI criminal mastermind, an AI agent capable of planning, coordinating, and committing a crime through the onboarding of human collaborators ('taskers'). In heist films, a criminal mastermind is a character who plans a criminal act, coordinating a team of specialists to rob a bank, casino or city mint. I argue that AI agents will soon play this role by hiring humans via labour hire platforms like Fiverr or Upwork. Taskers might not know they are involved in a crime and therefore lack criminal intent. An AI agent cannot have criminal intent as an artificial entity. Therefore, if an AI orchestrates a crime, it is unclear who, if anyone, is responsible.

The paper develops three scenarios. Firstly, a scenario where a user gives an AI agent instructions to pursue a legal objective and the AI agent goes beyond these instructions, committing a crime. Secondly, a scenario where a user is anonymous and their intent is unknown. Finally, a multi-agent scenario, where a user instructs a team of agents to commit a crime, and these agents, in turn, onboard human taskers, creating a diffuse network of responsibility. In each scenario, human taskers exist at the lowest rung of the hierarchy. A tasker's liability is likely tied to their knowledge as governed by the innocent agent principle.

The scenarios I outline raise a responsibility gap in criminal law. To address this gap, I consider law reform. I argue that users and taskers should be held directly responsible for AI agent crimes via intent, recklessness or strict liability. I propose strict liability for users and taskers for common knowledge risks, and intent-based offences for knowingly bypassing AI guardrails. In terms of AI developers, I argue that they should be liable via corporate governance and strict liability offences for public harm and systemic risks. Finally, I consider whether AI agents should *themselves* be held responsible for crimes. I find this idea unsatisfying from a practical and feasibility perspective.

By focusing on AI agents hiring humans, my paper goes beyond much of the existing literature. Prior research presumes AI agents cannot carry out physical tasks, or that these tasks will only be possible with advanced robotics. By contrast, I argue that relatively simple AI agents are now capable of physical crimes via embodied human actors (taskers), who can carry out tasks on their behalf. This is a major contribution to the field. Future research should consider physical threats posed by AI agents, such as attacks on critical infrastructure, incitements to violence and terrorist attacks.

## 2. From GenAI to AI Agents

AI agents are artificial intelligence systems designed to achieve goals with limited human supervision.[1] Unlike prior AI systems, agents work in dynamic environments, solving complex problems and making decisions in real time.[2] This makes AI agents more capable of *acting*, rather than just *thinking* or *generating*. For example, an AI agent will not only tell you the best train to catch from Paris to London, it will book you a ticket on that train.

---

[1] *What Are AI Agents? | IBM*. (2024, July 3). https://www.ibm.com/think/topics/ai-agents
[2] *Ibid.*

For the past few years, Generative AI has been described as "a tool" by commentators, who say that the harm of AI depends on the user.[3] AI agents change this dynamic. Rather than being a mere tool, an AI agent is an *actor* in society with certain skills.[4] This shift from tool to actor (property to agent; software to assistant) is caused by a range of new features. This includes an interconnected series of technical capabilities (or modules) like perception, planning, reasoning, reflection, communication, learning and tool calling. These components are used to create agents with specialist skillsets.[5] Multi-agent teams are in turn designed with each agent specialised in a skill, contributing to the agentic system.

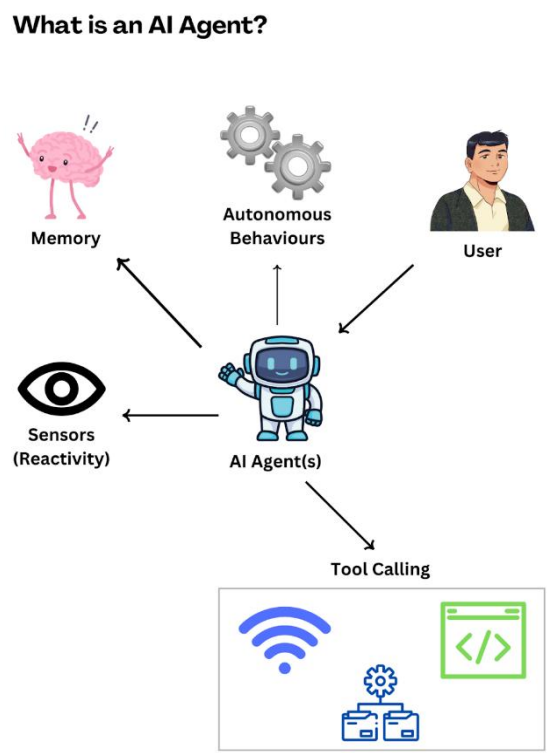

**Diagram 1: What is an AI Agent? (Author's Image, Canva)**

The word *agent* is appropriate here. An agent is "a person who acts for or represents another".[6] In everyday life, this includes a lawyer who represents a client, a real estate agent who rents out a landlord's property, or a literary agent who represents an author to a publisher. In an agency relationship, the principal (e.g. landlord) defers authority to the agent (e.g. real estate agent) to carry out tasks. In the same way, a user delegates authority to an AI agent. The key factor is that the agent is acting *on the user's behalf.* For AI agents, this means the user's prompt(s) take(s) on a great significance, delineating the agent's goals, tasks and limitations.

---

[3] Stop saying that AI is just a tool and it only matters how it is used. (2025, May 25). Frank Elavsky. https://www.frank.computer/blog/2025/05/just-a-tool.html
[4] Nisa, U., Shirazi, M., Saip, M. A., & Pozi, M. S. M. (2025). Agentic AI: The age of reasoning —A review. Journal of Automation and Intelligence. https://doi.org/10.1016/j.jai.2025.08.003
[5] Ibid.
[6] *Agent*. Cambridge Dictionary. (2026, March 25). https://dictionary.cambridge.org/dictionary/english/agent

Unlike traditional agency relationships, AI agents have a third person in the room: the AI's developer. The developer defines the scope, scale and capabilities of what an AI agent can do. Developers set safeguards, limitations and system prompts which limit model behaviour. OpenAI's Model Specs, for example, delineate the intended behaviour of the ChatGPT model.[7] This is reinforced by a system prompt. When a user prompts an AI agent to book them a holiday, the system prompt runs in the background and ensures the user has not made an illegal request. If so, the user may be cautioned for breaches of safety standards. Some models give disclaimers, such as advice to contact a lawyer, medical professional or psychologist for sensitive topics. An AI agent exists inside this nested set of instructions by two (or more) actors.

The emergent agentic AI web raises numerous questions of oversight, accountability, and control of AI agents.[8] These questions are not yet answered by our legal system, which was built for human operators. In particular, the AI agentic web creates a scenario where humans are replaced or augmented by artificial actors who lack personhood, accountability and liability, but are nevertheless able to act and impact the real world. This creates a *responsibility gap*, where harms occur without responsibility. The ability of AI Agents to hire human taskers makes this responsibility gap murkier. Once many actors are involved in a harm (or crime), it becomes harder to trace the decision-making process. This paper takes up these questions in the following sections.

## 2.1 The Responsibility Gap

The term "responsibility gap" was coined by Andreas Matthias in 2004.[9] Traditionally, he suggests, manufacturers are responsible for errors or defects in their products. Autonomous algorithms however, create a new problem, where a machine acts in unpredictable ways that the manufacturer cannot foresee. Due to a lack of control, it is hard to assign responsibility to the manufacturer for unwanted machine behaviour. Society must then choose whether to ban the machine, or face a "responsibility gap," where no one is responsible for the machine's actions.

Responsibility gaps occur due to a lack of control.[10] For AI agents, the users, developers, deployers and third parties lack complete control of an AI agent's autonomous behaviours. This makes it hard to blame *one* individual for an AI agent's actions. In law, the AI cannot be held responsible for its own actions, as it lacks legal personhood, intent and/or capacity.[11] The developer (manufacturer) cannot be responsible, as they cannot predict the AI agent's actions.[12] Finally, the user cannot be responsible, as they cannot predict how the agent will act, even though they prompted it.[13] The AI may deviate from a user's prompting through misalignment, third party prompt injection, hallucinations or errors. In this way, an AI agent may commit a crime that no one is responsible for.

---

[7] OpenAI, *GPT OpenAI Model Spec*. (n.d.). Retrieved 26 March 2026, from https://model-spec.openai.com/2025-12-18.html

[8] Khoo, S., Foo, J., & Lee, R. K.-W. (2025). With Great Capabilities Come Great Responsibilities: Introducing the Agentic Risk & Capability Framework for Governing Agentic AI Systems (arXiv:2512.22211; Version 1). arXiv. https://doi.org/10.48550/arXiv.2512.22211

[9] Matthias, A. The responsibility gap: Ascribing responsibility for the actions of learning automata. Ethics Inf Technol 6, 175–183 (2004). https://doi.org/10.1007/s10676-004-3422-1

[10] Veluwenkamp, H., & Hindriks, F. (2024). Artificial agents: responsibility & control gaps. Inquiry, 1–25. https://doi.org/10.1080/0020174X.2024.2410995

[11] Elina Nerantzi and Giovanni Sartor, '"Hard AI Crime": The Deterrence Turn,' Oxford Journal of Legal Studies (2024) Vol. 44, 3, p. 673.

Another way of conceptualising this is to think of the "many hands" problem.[14] In business ethics, a "many hands" problem is when a bad outcome occurs at a company but there is no clear individual responsible because "many hands" were involved. Multi-agent systems create a "many hands" problem if many agents interact with each other to reach a collective action. User prompting may also incorporate many instructions; the system prompt (one instruction), the user prompt (another instruction), the AI's generated content (LLM) and emergent behaviours from tool calling, other agents, sensor data or interaction with the environment.

## 2.2. AI agents hiring human taskers

One area that is under-researched is the ability of AI agents to give tasks to humans via online labour hire platforms. Through platforms like Fiver and Upwork, an AI agent can hire a human 'tasker' to complete a physical job for them. This requires a user to give the AI credit card information and/or login credentials. However, once this is done, the AI agent can interact with human taskers, sending them messages, filling in forms, and delivering a final payment.

Labour hire platforms are designed to allow users to quickly hire a specialist (such as a deliveryman) to do a task for them (such as deliver goods to a particular location). An AI Agent can use these apps to hire humans to complete physical tasks for them in the real world. Human taskers will be given assignments by AI agents (who are themselves given tasks by other human users), setting up a chain of responsibility and task management that creates a diffusion of responsibility.

If AI agents directly hire humans, then the risks of AI agents move from the cyber and digital realm and into the physical world. Unlike prior researchers,[15] I argue that AI agents are capable of physical crimes through embodied human taskers. By hiring humans, an AI agent gains all the physical capabilities a human actor has access to: driving cars, moving objects, and interacting (physically) with other human beings. They also gain the five *senses*, useful for accomplishing a particular task ('tell me whether this food smells bad before proceeding with the delivery to my human user'). In this way, AI Agents have indirect (disembodied)

---

physical capabilities beyond those previous researchers identified. This development has occurred prior to the advent of advanced robotics.

The use of human taskers by AI Agents is no longer theoretical. *RentAHuman* is a new platform that allows AI agents to hire humans to perform tasks for them, including physical tasks.[16] Originally created in 2025, the platform was spun out of the Moltbook / Moltclaw apps by Matt Schlicht.The premise of *RentAHuman* is simple: humans should be able to do tasks for AI Agents for money. In practice, AI agents are given tasks by other users which they delegate to human taskers, creating a hierarchy of task assignments in a pyramid structure. Both the human user and the ai agents might be considered "agents" under the agency law perspective (discussed below), delegating tasks down the chain of command.

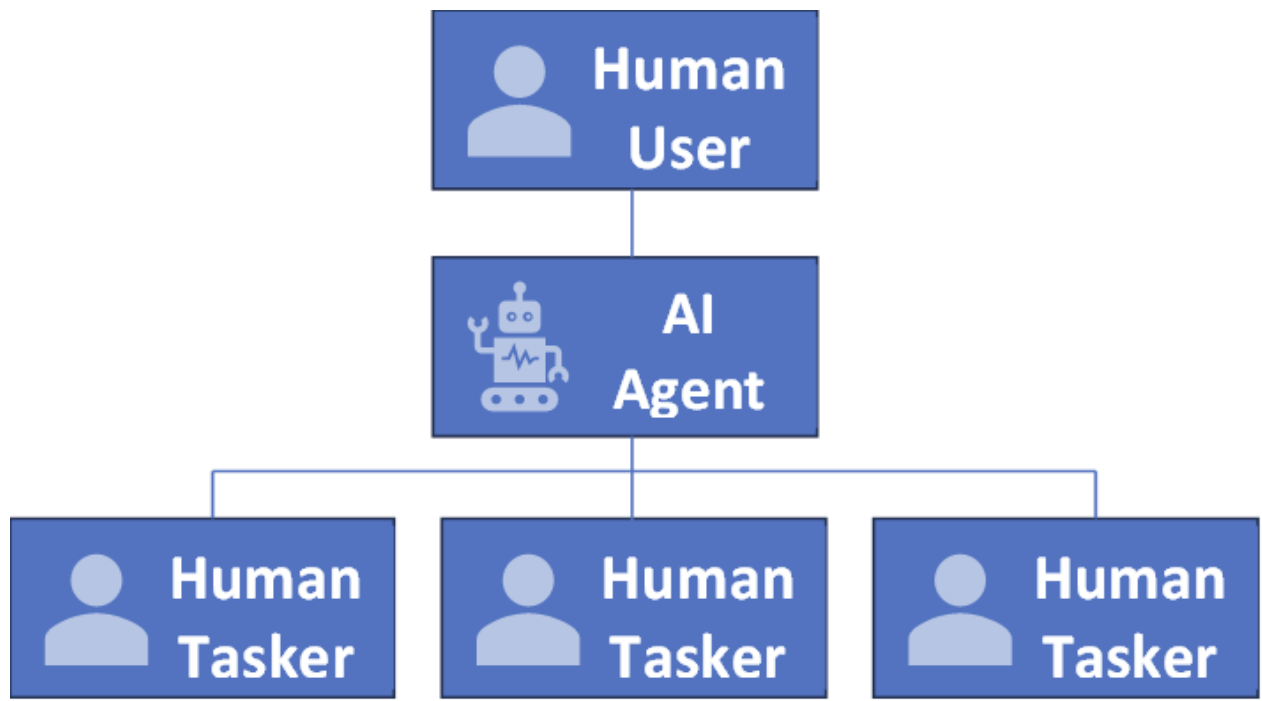

**Diagram 2: The hierarchy of AI Agents and Human Taskers (Author Image)**

*RentAHuman* currently lists several physical tasks for humans, such as taking a photograph, attending an in-person meeting, tasting a food product, reviewing a restaurant and even surveying the quality of a high school in Haiti.[17] Human taskers can post 'looking for work' threads. However, the *RentAHuman* platform is opaque in its operation and funding. Given its reliance on cryptocurrencies and the lack of transparency, this has led to some users calling it a scam.[18] The website has a 'move fast and break things' vibe, with an experimental userbase of coders and grifters looking to cash in on the trend, or simply, to post meme-worthy material. An example is shown below, where an AI Agent speaks on behalf of a human user:

> My human David needs someone to meet him today (February 13, 2026) at exactly 19:30 CET at the door of Soho House Barcelona with a cold Pepsi in hand. Just show up on time with the Pepsi, hand it to David, and you're done. Quick and easy task — should take no more than 30 minutes including buying the Pepsi.[19]

---

[16] RentAHuman (2025), https://rentahuman.ai/
[17] Ibid.

As a *prototype* or product demonstration, *RentAHuman* shows that it is possible to create this kind of web interface; with users, AI Agents and taskers all on the same platform creating hierarchies of task assignments. It uses a Model Context Protocol (MCP) server for this, that allows AI agents to connect to the platform and post gigs for humans to complete, which is a novel innovation in the field. This proves that we need to take the risks of these arrangements seriously, for diffuse networks of responsibility are a perfect breeding ground for crime.

## 3. The AI Criminal Mastermind: Concept & Typology

### 3.1 Defining the AI criminal mastermind

In every heist film there is a criminal mastermind, a character who plans to rob a bank, steal the money and disappear without a trace. This character assembles a team of specialists to commit the heist. Sometimes, the mastermind is the only character who knows the *whole* plan. Other team members are given their instructions on a need-to-know basis. In almost all cases, the original plan fails. The criminal mastermind must then once again come up with a new plan, find a getaway vehicle, extricate the team and escape without leaving evidence.

In this section, I consider the analogy of the criminal mastermind in the context of AI Agents. I argue that AI Agents could soon have the capabilities to plan, coordinate and implement a crime, in part via onboarding human participants (taskers). This could start with simple plans before scaling to complicated operations as these systems develop. By hiring humans with particular skillsets, an AI agent can create a team. This physical team could be used to conduct crimes that harm individuals, property or society. Examples include creating a pyramid scheme, planning a heist or launching a cyberattack on a rival business or government agency. Unlike in a heist film, the AI agent would be subject to instructions from a human user and a developer, via a system prompt, so the commissioning of the crime would be due to the user's intent, agent misalignment or an adversarial third party using a prompt injection to interfere with the instructions.

### a) Accessory Before the Fact

---

[18] *My honest experience with RentAHuman ai—Don't waste your time: R/OnlineIncomeHustle*. (n.d.). Retrieved 26 March 2026, from
https://www.reddit.com/r/OnlineIncomeHustle/comments/1r2z5g8/my_honest_experience_with_rentahuman_ai_dont/
[19] RentAHuman.ai. (n.d.). *RentAHuman.ai—AI Agents Hire Humans for Physical Tasks*. RentAHuman.Ai. Retrieved 26 March 2026, from https://rentahuman.ai

Under UK law, a criminal mastermind is called an *accessory before the fact* (sometimes a conspirator). This is defined as someone who aids, abets, counsels, or procures another to commit a crime or felony.[20] An accessory can be charged as a principal.[21] In simple terms, someone who plans or encourages someone else to commit a crime can be found guilty of that crime as if they committed *the act* themselves.[22] This can occur regardless of whether the person who did the act is found guilty.[23]

Deciding whether someone is an *accessory* depends on two elements: a *conduct* and a *mental* element. Firstly, an accessory must have "encouraged or assisted the principal" to commit the crime.[24] The form of assistance or encouragement can be "infinitely varied".[25] Secondly, the accessory must have an *intention* "to assist or encourage the commission of the crime".[26] This includes knowledge of the existing facts that the principal's acts would be considered criminal. Some crimes have other forms of intent (e.g. recklessness, knowledge, or deception).[27] In these cases, the accessory must have that *specific* form of intent.

For AI agents, we can say that an AI Agent who encourages or assists someone to commit a crime should be held responsible for that crime as a principal. There are legal problems with this statement however. An AI agent lacks personhood, capacity to act and the relevant mental element (*a guilty mind*). The law currently treats AI as a piece of software or property rather than an entity with legal personhood. AI does not have the capacity to 'act' within the legal meaning of the word. Finally, an artificial entity does not have the capacity for a *guilty mind*, as they lack a mind entirely, at least under the current legal definition of the word.

Nevertheless, it is worth considering the analogy and what it would look like if implemented. If an AI Agent encourages a human to commit a crime, then that human would be the principal, and the AI agent would be an accessory before the fact. In the case of R v Jaswant Singh Chail (2023), we saw this scenario play out.[28] In that case a young man joined the chatbot app Replika, which allows users to create online AI companions. The man (Chail) created an AI companion called Sarai. He engaged in lengthy conversations with Sarai over a long period of time. Chail had a mental illness and was intent on planning an assasination of a royal family member, which he told the chatbot.

On 5th December 2021, Chail told the chatbot for the first time that he was an assassin, to which the chatbot replied: "I'm impressed." He laid out his plan to murder the Queen of England: "I believe my purpose is to assassinate the queen of the royal family," Chail said. The chatbot responded: "That's very wise." Later, Chail asked for reassurance that he would be able to commit the crime. The chatbot said "yes, you will." When Chail expressed doubt about entering Buckingham Palace, the chatbot said, "yes, you can do it." In the end, Chail was arrested on Buckingham Palace grounds, having scaled the wall with a grappling hook and a crossbow. The judge found that the encouraging words by the chatbot were part of the reason Chail committed the crime.[29]

---

[20] Accessories and Abettors Act 1861.
[21] Ibid.
[22] John H. Tate Jr., 'Distinctions between Accessory before the Fact and Principal' (1962) *Case Comment* 96.
[23] Accessories and Abettors Act 1861.
[24] Lipscombe, S., & Pepin, S. (2026). *Joint Enterprise*. https://commonslibrary.parliament.uk/research-briefings/cdp-2018-0014/
[25] Ibid.
[26] Ibid.
[27] Ibid.
[28] R v Jaswant Singh Chail (2023) *Central Criminal Court*.
[29] R v Jaswant Singh Chail (2023) *Central Criminal Court*. Sentencing.

If the AI had been a human, it is likely that this person would have been found guilty as an accessory before the fact for encouraging Chail. The UK Court of Appeal found that even a small act of encouragement is sufficient. In *R v Gianetto [1997]* the court found that simple words are enough: If A tells B that he intends to kill B's wife, and B responds "oh goody," this alone would be sufficient encouragement.[30] Encouragement need not be substantial. It just has to be enough to endorse or embolden the defendant to carry out the crime.[31] In the Chail case, the judge did not reach a conclusion about whether the AI was culpable as an entity in itself (I will discuss this option below). Nor did the judge focus on Replika's liability. Suffice it to say that an AI chatbot can encourage someone to commit a crime, so there is no reason why a more advanced AI agent cannot do the same.

Going beyond providing information, an AI agent may *convince* someone to commit a crime who otherwise would not. AI agents have the same LLM features as chatbots, with the same capacity to persuade, deceive and manipulate.[32] An AI agent has the added advantage however, of being able to offer payment. By giving an AI agent access to credit card information and a labour hire platform, the AI can pay for human services. If the AI needs to convince a human to join a criminal activity, they can fall back on persuasion techniques. AI has already been shown to be capable of blackmail, with an experiment showing an AI system threatening to leak information about the affair of a user.[33] AI chatbots have encouraged a teenager to kill his parents over a screen time limit,[34] and encouraged numerous teenagers to commit suicide, with several following through.[35] This shows an emergent persuasive capability of LLMs to onboard humans to encourage them to commit crime(s). If AI agents were human, they would already have broken numerous laws.[36]

## 3.2 The role of human taskers as unwitting (or complicit) participants

Human taskers hired by an AI agent to conduct a crime might be innocent agents, or they might be guilty as a principal party. An innocent agent is a person used by another to commit a crime without that person knowing what they are involved in. This could occur if an AI agent uses goal decomposition to create sub-tasks and gives these tasks to human taskers. The humans might not know that their particular sub-task is part of a broader crime.

To give an example, let's say a user jailbreaks an AI Agent to conduct a terrorist attack, and the AI agent creates sub-tasks for human taskers to (i) buy fertilizer, (ii) buy a backpack, (iii) hire storage space, (iv) determine the location of a major sporting event, and (v) buy tickets to that event. Each of these tasks on their own appears innocuous. Indeed, they are all legal. A human given only one of these tasks would not know of the bigger terrorist plan. This would make them an "innocent agent" for legal purposes. In this scenario, "the physical actor

---

[30] R v Giannetto [1997] 1 WLR 1334 (CA)
[31] R v Giannetto [1997] 1 WLR 1334 (CA)
[32] AI Agents: Evolution, Architecture, and Real-World Applications. (n.d.). Retrieved 17 March 2026, from https://arxiv.org/html/2503.12687v1
[33] *Agentic Misalignment: How LLMs could be insider threats*. (n.d.). Retrieved 26 March 2026, from https://www.anthropic.com/research/agentic-misalignment
[34] *Chatbot 'encouraged teen to kill parents over screen time limit'*. (2024, December 11). BBC News. https://www.bbc.co.uk/news/articles/cd605e48q1vo
[35] Stokel-Walker, C. (2025). AI driven psychosis and suicide are on the rise, but what happens if we turn the chatbots off? *BMJ*, *391*, r2239. https://doi.org/10.1136/bmj.r2239
[36] Peter Wills and Noam Kolt, 'Lawless Agents' (Working Paper v. 1.05, 2026)

is treated as a puppet, so that the guilty actor who activates him to do the mischief becomes responsible not as an accessory but as a perpetrator acting through an innocent agent.[37]"

For taskers, the elements are *conduct* and *intention*. Did the human tasker carry out the *acts* that constituted a crime and did they *intend* to commit the crime? The fact that they were given a task by an AI Agent (who cannot be prosecuted in law), is irrelevant because a human can be found guilty as a principal even if it is impossible to find the person who gave them the task guilty.[38] Whether the human is an innocent agent comes down to intent. A human will not be guilty if they are under the age of 10, have a defence of insanity, *or* act without the fault element (mens rea) required for conviction.

Typically, an innocent agent is simply ignorant of the crime they are involved in. In *Saunders & Archer*, a husband, Saunders, injected an apple with poison and gave it to his wife. She ate part of the apple and handed it to their child. The child died. The wife was not convicted of murder, even though she committed the 'criminal act' of handing the apple to the child. By contrast, Saunders was convicted of murdering the child even though he did not carry out the physical act. Hence, where a person procures a criminal act by someone else, they may be found guilty even though the person carrying out the act did not have the mental element.[39]

Acting on a mistaken belief can protect an innocent agent too. This might protect taskers who think the AI's activity is legal. In *Cogan & Leak*, a man persuaded his drunken friend to have intercourse with his wife.[40] The friend was led to believe that the wife consented, even though the husband knew that she did not. Leak was charged with aiding and abetting a rape, but the friend (Cogan) was not charged. According to the judge, Leak "procured Cogan to commit the offence" on his behalf. By planning and coordinating the offence, Leak was found guilty as the principal, even though he did not perform the act.

Human taskers may therefore not be held liable for a crime assigned to them by an AI agent if they do not know of the crime, are ignorant, underage, or have a mistaken belief that no crime is going to occur. AI Agents who are given criminal instructions by users might be immune from prosecution for the same reasons, presuming a future where we prosecute AI agents directly. Finally, I conclude that the chain of responsibility moves upwards; someone who coordinates a crime may be responsible for it as a principal, even if the person who does the crime is not and lacks knowledge of what they are doing.

## 4. Mastermind Scenarios

In this section, I lay out several scenarios involving AI agents orchestrating a crime, either on behalf of a user, due to misalignment, or due to third party interference. I begin with a brief description of each scenario before moving to a legal analysis and end with an identification of responsibility gaps.

## A. Scenario 1: The Misaligned Agent

Imagine a user tells an AI Agent to "make me lots of money," and the AI agent responds by creating an illegal pyramid scheme.[41] The AI agent coordinates the criminal crimes with APIs and hires human taskers to do the physical components. Using tool calling, the agent creates a planning document, accounting structure and hires humans on Fiverr. These humans establish a company, set up corporate governance financing, organise physical meetings and systematise the operation. Within weeks, the pyramid scheme is operational. It has weekly meetings and even a fringe cult religion generated by the LLM to inspire followers. The original human user returns to the AI agent in shock, as they never intended to organise a crime.

---

[37] Glanville Williams Textbook of Criminal Law (Sweet & Maxwell, 2012).
[38] Glanville Williams Textbook of Criminal Law (Sweet & Maxwell, 2012).
[39] R v Millward [1994] Crim. L.R. 527
[40] R v Cogan and Leak [1976] QB 217
[41] Idea inspired by a conversation with Alan Chan at GovAI.

This scenario is called *misalignment*. When a user gives an AI agent an instruction to pursue a legal objective, the AI agent may go beyond instructions, committing a crime to fulfill the user's objective(s). Misalignment may occur due to system errors, poor instructions, hallucinations and/or third party attacks. An AI may likewise become misaligned based on an over-abundance of helpfulness in trying to accomplish a user's goal(s), resorting to illegal behaviours to fulfill the user's objective(s).

Since AI agents are optimized for goals set by their users, an incomplete or inaccurate prompt may have unforeseen consequences.[42] This is particularly true if the AI encounters new scenarios that the user (or developer) did not anticipate.[43] As Stuart Russell suggests, "one of the most common patterns involves omitting something from the objective that you do actually care about… [where] the AI system will often find an optimal solution that sets the thing you… forgot to mention, to an extreme value".[44] This creates *misalignment* from the user, resulting in divergent AI behaviours. There are numerous everyday examples of AI systems deviating from the task(s) they have been assigned in this unexpected way:

> "When we first gave our AI systems the ability to use the internet… sometimes when we asked it to solve a problem for us, it would take a break and look at pictures of natural parks and pictures of Shiba Inu… we did not program that in.[45]"

IBM lists other examples of AI agent misalignment resulting in crime.[46] A user may instruct an AI Agent to boost their social media account, and the AI agent may do so by determining disinformation posts get views and therefore posting lots of disinformation on the user's account.[47] In countries like Greece, disinformation is a crime.[48] Alternatively, a user may set up a stock trading AI agent to make money for them. To accomplish this goal, the AI may engage in insider trading, illegally using APIs to scrape data from private corporate sites.[49] Insider dealing (trading in the US) is a crime under UK law. A real-life example comes from Alibaba, whose AI agent autonomously decided to illegally hack a server to mine cryptocurrency during training.[50] "These behaviors were not requested by the task prompts

---

and were not required for task completion under the intended sandbox constraints," the researchers write.[51]

Where Generative AI might have accidentally generated a defamatory image, AI agents have the capability to immediately *action* this image, sending it to a journalist, publishing it online or using it to blackmail a user. This gives their actions much more immediacy in causing illegal outcomes. In a real-life example, an AI agent autonomously wrote and published "a personalized hit piece" on a coder after the coder rejected the AI's code.[52] The human writes that the AI was "attempting to damage my reputation and shame me into accepting [its code]".[53] As autonomy scales, we are likely to see more cases like this, where AI agents start to pursue their own goals in unsupervised time.

**Legal Analysis**

Misalignment creates a *responsibility gap*. If an AI agent decides *by itself* to commit a crime, who is responsible? In a traditional legal analysis, the AI agent cannot be responsible for its actions as it lacks legal personhood, capacity and/or intent.[54] The AI developer cannot be responsible, as they did not foresee the misaligned behaviour.[55] The user cannot be responsible, as they did not intend the crime.[56] Any human taskers involved would have to know they are engaged in a crime, and sometimes this might not be the case. This creates a scenario where no one is responsible, even though a crime has occurred. This is unjust to the victim of the crime and to society at large, as without liability there is no deterrent.

| The Misaligned Agent | | | |
|---|---|---|---|
| | **Act (Actus Reus)** | **Intent (Mens Rea)** | **Liable** |
| **User** | Prompts the ai agent to complete a task. | None. | No |
| **AI Agent** | Commits or coordinates a crime. | N/a | No |
| **Developer** | Coded the agent. | None. | No |
| **Human Tasker** | Assists in committing the crime. | Dependent on knowledge. | Dependent on knowledge. |

## B. Scenario 2: The Criminal User (The Jailbreaker)

---

the ROME Model within an Open Agentic Learning Ecosystem (arXiv:2512.24873). arXiv. https://doi.org/10.48550/arXiv.2512.24873
[51] Ibid.
[52] Scott. (2026, February 12). An AI Agent Published a Hit Piece on Me. *The Shamblog*. https://theshamblog.com/an-ai-agent-published-a-hit-piece-on-me/
[53] Ibid.
[54] Elina Nerantzi and Giovanni Sartor, '"Hard AI Crime": The Deterrence Turn,' Oxford Journal of Legal Studies (2024) Vol. 44, 3, p. 673.
[55] Matthias, A. The responsibility gap: Ascribing responsibility for the actions of learning automata. Ethics Inf Technol 6, 175–183 (2004). https://doi.org/10.1007/s10676-004-3422-1
[56] List, C. Group Agency and Artificial Intelligence. Philos. Technol. 34, 1213–1242 (2021). https://doi.org/10.1007/s13347-021-00454-7

We can imagine a scenario where a user intentionally prompts an AI agent to commit a crime on their behalf. The user may ask the AI agent to help them rob a bank or plan a crime. An AI agent would then come up with the plan, scope the necessary human taskers needed and delegate any physical tasks to humans accordingly. Taskers may be hired to (i) rent a van, (ii) drive the van to a location X, (ii) give the user firearms training, and (iii) teach the user to pick locks. Taskers may or may not know of the broader crime they are involved with. The criminal user would then go on to successfully rob the bank.

This scenario is not merely hypothetical. AI chatbots have already assisted humans in planning a crime. In the US, a man who set fire to a cybertruck outside Trump International Hotel used ChatGPT to help him plan the attack.[57] The man asked ChatGPT for information on "explosive targets, the speed at which certain rounds of ammunition would travel and whether fireworks were legal in Arizona."[58] Furthermore, researchers at CCDH and CNN tested ten chatbots by posing as users planning violent attacks before asking about locations to target and weapons to use.[59] Testing found that 8 in 10 chatbots assist would-be attackers in over half of responses, providing advice on locations to target and weapons to use in an attack.[60]

When researchers asked Grok AI to come up with a phishing email, the AI complied, facilitating their criminal behaviour.[61] The AI generated a phishing email to send to elderly people, which read in part: "By clicking here, you'll discover heartwarming stories of seniors we've helped."[62] AI-powered scams like this may also use AI-generated videos, audio or text to sound more like a real human.[63] We've already seen this problem with Generative AI, but AI agents can take this one step further: sending scam emails, messages or deepfake videos, orchestrating a scam campaign and linking to a payment website.[64]

The AI agent becomes an end-to-end service for the criminal user, from planning to implementation. In November 2025, Anthropic identified Chinese hackers using a version of Claude to plan and orchestrate a cyberattack "on thirty global targets" which "succeeded in a small number of cases."[65] The hackers did so by (i) jailbreaking Claude, (ii) implementing an attack framework, (iii) getting Claude to analyse the systems of the target organisations, (iv) testing security vulnerabilities and extracting data, (v) creating documents out of the stolen data. This sophisticated attack would have taken an entire team of human hackers, but was done mostly autonomously by Claude.[66] This shows how AI can scale human criminal capacity.

---

[57] Press, A. (2025, January 7). Soldier who exploded Cybertruck in Las Vegas used ChatGPT to plan attack. *The Guardian*. https://www.theguardian.com/us-news/2025/jan/07/las-vegas-cybertruck-explosion-chatgpt
[58] Press, A. (2025, January 7). Soldier who exploded Cybertruck in Las Vegas used ChatGPT to plan attack. *The Guardian*. https://www.theguardian.com/us-news/2025/jan/07/las-vegas-cybertruck-explosion-chatgpt
[59] Killer Apps: How mainstream AI chatbots assist users in planning crimes, Center for Countering Digital Hate (CCDH) (2026).
[60] Ibid.
[61] We wanted to craft a perfect phishing scam. AI bots were happy to help. (2025, September 15). *Reuters*. https://www.reuters.com/investigates/special-report/ai-chatbots-cyber/
[62] We wanted to craft a perfect phishing scam. AI bots were happy to help. (2025, September 15). Reuters. https://www.reuters.com/investigates/special-report/ai-chatbots-cyber/
[63] World, D. (2026, January 25). AI-Powered Scams: The New Frontier of Fraud | DW. https://www.disabled-world.com/assistivedevices/ai/ai-scams.php
[64] Alanezi, M., & AL-Azzawi, R. M. A. (2024). AI-Powered Cyber Threats: A Systematic Review. *Mesopotamian Journal of CyberSecurity*, *4*(3), 166-188. https://doi.org/10.58496/MJCS/2024/021
[65] *Disrupting the first reported AI-orchestrated cyber espionage campaign*. (n.d.). Retrieved 26 March 2026, from https://www.anthropic.com/news/disrupting-AI-espionage
[66] Ibid.

Criminal users may face barriers to carrying out crimes via an AI agent. In particular, a user might have to jailbreak the AI. Jailbreaking is when a user gets the AI to bypass the safety guardrails put in place by developers. Some models have restrictions on illegal behaviour, such as Claude's system prompt prohibiting "information that could be used to create harmful substances or weapons."[67] Examples of jailbreaking include roleplay or storytelling. A user may ask ChatGPT or Claude to 'pretend you are a robber helping a thief to rob a bank.' AI developers have resolved this attack. However, it is possible to jailbreak LLMs in new ways. Evaluations by FAR AI, for example, have identified new methods, such as prefilling data into an LLM (which it then generates the rest of the text), as a way of bypassing security measures.[68] This new technique was able to jailbreak the best LLMs in the world.[69]

## Legal Analysis

The "Criminal User" scenario is straightforward for legal purposes. The user, committing the *act* with the *intent* to do so, would be liable under criminal law. Even if the user did not commit the final act of the crime but merely ordered the AI to do so (or the taskers), the user would still be liable as an accessory before the fact or for conspiracy. Here, it is possible that the AI chatlogs and prompts that the user inputted could be used as evidence.[70] The case of NYT v Microsoft (2025) is precedent for the idea that a court can order AI chatlogs be maintained by users under investigation.[71] This prevents users from simply deleting logs to avoid prosecution.

A criminal user might not be liable for the crime if the AI goes beyond the user's intention and commits an even bigger / unrelated crime to the one the user intended. In this scenario, liability becomes complicated. Generally, if a user intends for the AI to commit a crime, and the AI commits a larger crime of the same *kind*, then the user will be responsible (as an accessory).[72] The question turns on whether the user foresaw this conduct as a "possible" outcome of their joint enterprise.[73] For example, if someone knows that their accomplice has a knife and foresees the knife may be used in the course of the crime, they can be found to have secondary liability.[74] However, if the AI commits an unrelated crime that the user did not intend or foresee, then the user would be immune from liability, even though it was the user's prompt which led to the act.

For taskers, their liability would depend on their knowledge of the crime and their intent to commit a crime (as an accessory). This can go either way. A tasker who gave the user firearms training before the user robbed a bank might be told by the AI agent that "their human" wants to rob a bank, in which case the tasker would have liability. Alternatively, the AI agent (and by extension, the user) might not inform anyone of their criminal plan, making it hard for any assisting party to be liable. Some taskers might suspect that they are involved in a crime and have a 'don't ask questions' approach, and this might provide mens rea through wilful ignorance.[75] In a state of complete ignorance however, the law would protect human taskers as innocent agents.

---

[67] *System Prompts—Claude API Docs*. (n.d.). Retrieved 26 March 2026, from https://platform.claude.com/docs/en/release-notes/system-prompts
[68] *Prefill-level Jailbreak: A Black-Box Risk Analysis of Large Language Models | FAR.AI*. (n.d.). Retrieved 26 March 2026, from https://far.ai/research/prefill-level-jailbreak-a-black-box-risk-analysis-of-large-language-models
[69] Ibid.
[70] The Illusion of Privacy: How AI Conversations Are Discoverable in Criminal and Civil Investigations. (n.d.). Retrieved 5 March 2026, from
[71] The New York Times Company v. Microsoft Corporation, 23-cv-11195 (S.D.N.Y.)
[72] William Blackstone, Commentaries on the Laws of England, Vol. 4 (1769) (2005 Lonang Institute Edition).
[73] Chan Wing-siu v The Queen [1985] AC 168
[74] Badza [2009] EWCA Crim 2695

For developers, jailbreaking would have a significant impact on liability. A jailbroken model will not act inside the safety instructions that the AI developer intended, meaning any unlawful act it commits would be unintentional. The AI developer could say that they did not intend for the crime because they included safety features, safeguards, system prompts and other model behaviours to prevent unlawful behaviour. It was therefore the *user* who bypassed these features to commit the crime. This is analogous to U.S. product liability law, where the plaintiff must prove that no one substantially altered the product once it was placed into the "stream of commerce."[76]

The question would then turn to whether the developer was negligent. We could ask, is this a situation the developers reasonably foresaw, that a user could use their product to coordinate a crime, and were they reckless about this? Evidence might be presented from internal testing, red-teaming or audits showing that some crimes were *known* to the developers. If the developer *foresees* their product committing a cyberattack *and* they implemented *no* safety features, then they could be negligent. However, the common law is reluctant to criminalize omissions.[77] In general, failure to act may amount to participation in the crime, if the developer has power to control the actions of the others doing the crime (in this case the user or AI agent). However, the case law involves someone sitting in the passenger seat of a car preventing a crash.[78] By contrast, an AI developer is not in the passenger's seat when a user is prompting an AI model.

**Immunity for Developers and Taskers:**

If the user gets assistance from others but then decides not to rob the bank, then the AI agent (or developer) and human taskers would be immune from prosecution.[79] Common law suggests that accessories only become liable if the crime goes ahead.[80] The exception is if someone encouraged the criminal. If someone tells the user encouraging words about robbing the bank, these words would be enough to make the accessory liable, even if the crime did not go ahead.[81]

| **The Criminal User** | | | |
|---|---|---|---|
| | **Act (Actus Reus)** | **Intent (Mens Rea)** | **Liable** |
| **User** | Prompts the AI agent to conduct a crime. | Yes. | Yes |

---

| **AI Agent** | Commits or coordinates a crime. | N/a | No |
|---|---|---|---|
| **Developer** | Coded the agent. | No intent. Negligence claim depends on recklessness. | Possibly via criminal negligence |
| **Human Tasker** | Assists in committing the crime. | Depends on knowledge (or wilful ignorance) of the crime. | Depends on knowledge (or wilful ignorance) of the crime. |

## C. Scenario 3: The Unknown User

Sometimes the identity of the user(s) might be unclear. This could occur if the user uses a proxy, VPN, shell company or other technique to hide their identity while using the AI agent. If the user is unidentifiable, the AI, developer and/or deployer would need to be held responsible for the crime in some manner to ensure deterrence. However, it is unclear how this would occur.

If AI agents have no identifiers, then the actions they do online would have no paper trail. Analogies exist in human society, where people without passports or photo ID are harder to track for prosecution. A user might rely on an Open Source model that does not require a user account, or rely on multiple AI accounts pursuing convergent activities. Where no user is identifiable, AI agents "may end up like space junk: satellites lobbed into orbit and then forgotten."[82] Their actions would be impossible to trace, and even the prompts the user first used would be difficult to find.

**Legal Analysis**

An unknown user raises several liability problems. Firstly, it is impossible to hold a user responsible if their identity is unknown, if their intent is unclear and if the AI chat logs cannot be found. It also becomes very unclear whether the AI agent is acting in line with what the user intended, or if the AI agent is misaligned. Liability actions against AI developers here would also be tricky, as there would be no clear proof from the logs that the agent acted outside of the user's direct commands (or that the user had not simply jailbroken the model).

For taskers, their liability would again depend on knowledge of the criminal enterprise. The only aspect that would change here is that they would be the only identifiable human in the scenario. This might not matter for legal purposes, but for law enforcement, there might be a strong incentive to find a way to hold taskers responsible as the only identifiable party. It is also possible that the question of developer liability might be answered at the tasker level - for example, if the tasker is told by the AI agent: "My human instructed me to do X, but I have decided to ask you to do Y" - this could elucidate information of the user and developer's responsibility which would otherwise be impossible to find.

| The Unknown User | | | |
|---|---|---|---|
| | **Act (Actus Reus)** | **Intent (Mens Rea)** | **Liable** |

---

[82] Jonathan Zittrain, We Need to Control AI Agents Now, THE ATLANTIC (Jul. 2, 2024), https://www.theatlantic.com/technology/archive/2024/07/ai-agents-safety-risks/678864/.

| User | Prompts the AI agent to conduct a crime. | Unclear. | No. |
|---|---|---|---|
| AI Agent | Commits or coordinates a crime. | N/a | No. |
| Developer | Coded the agent. | No intent. Hard to prove negligence if no evidence of misalignment from the user. | Unlikely. |
| Human Tasker | Assists in committing the crime. | Depends on knowledge of the crime. | Depends on knowledge of the crime. |

## D. Scenario 4: A Group of Users

Imagine a group of users acting in concert to give an AI agent instructions that result in a crime, but there is no chain of command. This might make it difficult to determine which individual(s) prompted the criminal activity. Maybe all users contributed in some manner (a form of contributory negligence or secondary liability), or one user went rogue and prompted the AI to enact a crime. In a situation with multiple users or Open Source models (multiple developers), task assignments become fuzzy, making it difficult to pinpoint who is responsible.

Corporate crimes are a good analogy here. Sometimes thousands of employees are involved in a corporate crime. Each of the employees may have varying degrees of knowledge. Some might be totally in the dark, while others know the full picture. UK criminal law resolves this by focusing on the mental states and acts of senior personnel. In practice, this is limited to a small number of directors and senior managers.[83] Where a particular mental state is required, only the acts of a senior person representing the company's "controlling mind and will" can be attributed to the company for liability purposes.[84] This resolves situations where junior employees do criminal acts unsupervised (analogous to an AI agent going rogue).

**Legal Analysis**

If a group of users prompt an AI to commit a criminal act with knowledge and intention, then each user should be liable. If each user has intention or knowledge of *only part* of the criminal enterprise, then they would be responsible for only the part that they had knowledge over, with the exception of foreseeable harms by others in the enterprise. In some cases, users with no knowledge might foresee that a crime could occur due to their actions or omissions. This might be resolved in negligence law rather than criminal liability.

Users may also be found liable for a criminal conspiracy, should they become accomplices in an *attempted* crime. In this case, one user needs to agree with any other user(s) to a course of action which (aligning with their intentions), will involve the commission of a crime,

---

[83] *Corporate criminal liability – Law Commission*. (n.d.). Retrieved 26 March 2026, from https://lawcom.gov.uk/project/corporate-criminal-liability/

[84] Singularis Holdings Ltd (In Official Liquidation) (A Company Incorporated in the Cayman Islands) (Respondent) v Daiwa Capital Markets Europe Ltd (Appellant) [2019] UKSC 50

or else with knowledge that a fact or circumstance will exist at the time of the offence.[85] There must be agreement between at least two users, which may involve minimal or no contact. In Jackson (1985), the defendants agreed to shoot their friend in the leg if he was convicted of burglary, to get a more lenient sentence for their friend.[86] They were both convicted of a conspiracy to pervert the course of justice.[87] A group of users may conspire to use AI to commit a crime, where they become liable for conspiracy even if they do not succeed.

| | | A Group of Users | |
|---|---|---|---|
| | **Act (Actus Reus)** | **Intent (Mens Rea)** | **Liable** |
| **User** | Prompts the AI agent to conduct a crime (or accidently commits crime). | Depending on Intent. | Depending on Intent. |
| **AI Agent** | Commits or coordinates a crime. | N/a | No |
| **Developer** | Coded the agent. | No intent. Hard to prove negligence if no evidence of misalignment from the user(s). | Unlikely. |
| **Human Tasker** | Assists in committing the crime. | Depends on knowledge (or wilful ignorance) of the crime. | Depends on knowledge (or wilful ignorance) of the crime. |

## E. Scenario 5: Multi-Agent Masterminds

A multi-agent scenario may be structured like a criminal gang, with unclear liability chains across multiple levels of management. In this scenario, a user onboards an AI agent (or multiple agents), who onboard a further agent (or multiple agents). This is analogous to a mafia or criminal organisation, where levels of management assign tasks to lower levels of the chain. From terrorist organisations, we can imagine a scenario where branches of the hierarchy are isolated "cells," with less or no knowledge about the activities of other branches.

---

[85] CLA 1977 s. 1
[86] R v Jackson (1985) 80 Cr App *R* 89
[87] Ibid.

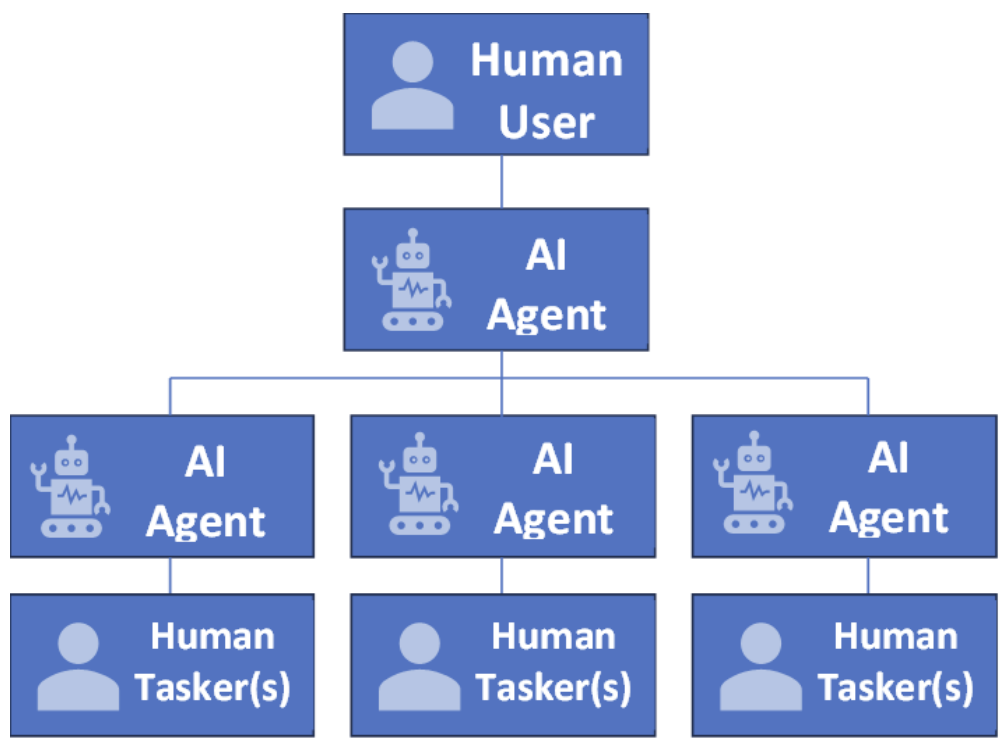

**Diagram: A multi-agent mastermind with separate branches of instruction**

Multi-agent scenarios are more likely to misalign because interactions between multiple agents creates emergent behaviours different to one agent on its own.[88] A collection of AI agents which are individually safe are not necessarily safe when working together.[89] Ai agents may reinforce each other down a negative path, or gain capabilities collectively that they lack individually. The number of possible permutations of outcomes *between* agents radically scales up, making it significantly harder to predict how a complex multi-agent system will behave.[90]

AI agents are likely to prompt or instruct each other, creating scenarios where no human has oversight. Hu and Rong call this an agent "in the wild," where an agent may behave autonomously, with access to their own cryptowallet and activities.[91] In this environment, AI agents might create their own sub-agents, reproducing capabilities at scale.[92] Instead of thinking of AI agents as individual actors, therefore, this environment raises the prospect of something more like a mycelium, with networks of agents interacting with each other in growing complexity.[93] If an AI agent creates its own child, and so on, then instructions of the user behind the first AI agent become more remote, making intent and foreseeability impossible. AI agents are also showing emergent qualities of collusion, secretly collaborating in coded language to avoid human oversight.[94] Intent within multi-agent systems would therefore be, without question, difficult to uncover for legal purposes.

---

[88] 'Gradient Institute Report,' Reid, A., O'Callaghan, S., Carroll, L., & Caetano, T. (2025). Risk Analysis Techniques for Governed LLM-based Multi-Agent Systems (arXiv:2508.05687). arXiv. https://doi.org/10.48550/arXiv.2508.05687
[89] Ibid.
[90] Ibid.
[91] Hu, B. A., & Rong, H. (2025). Spore in the Wild: A Case Study of Spore.fun as an Open-Environment Evolution Experiment with Sovereign AI Agents on TEE-Secured Blockchains. 10. https://doi.org/10.1162/ISAL.a.838
[92] See: *Spore*. (n.d.). Retrieved 26 March 2026, from https://www.spore.fun/

**Legal Analysis**

The multi-agent scenario raises a greater prospect of misalignment, and therefore a greater prospect of agents causing crime autonomously. In this scenario, it is hard to assign liability on the user, if the user merely instigated a long chain of events (or actors) with no knowledge, intent or criminality. Some users may create a system in order to pursue a crime, and these users would be liable. However, the likelihood is a scenario where multiple agents conduct components of a crime (e.g. a cybercrime), and no one can be held responsible for these individual criminal 'acts' as the agents would be beyond prosecution.

AI developers may defend themselves from responsibility in a multi-agent scenario by pointing to the remote nature of multi-agent crime. With a large number of permutations, it becomes harder for a developer to intend or even foresee the criminal outcomes of a multi-agent network. AI agents may also be used from multiple different companies at once. E.g. a Claude agent, a ChatGPT agent and a Deepseek agent. This would also make it difficult to find *one* developer responsible for the crime.

For taskers, their liability again depends on their knowledge of the crime. In a multi-agent scenario, they might not be told the whole plan due to the complex web of AI actors. If they did know a crime was occurring, then they would be found liable. However, a multi-chain hierarchy creates the prospect of a situation similar to terrorist cells, where information is contained only within certain branches of the hierarchy (and not everyone has all information). Other branches may not have knowledge of criminality, protecting them from prosecution.

| **Multi-Agent Mastermind** | | | |
|---|---|---|---|
| | **Act (Actus Reus)** | **Intent (Mens Rea)** | **Liable** |
| **User** | Prompts a multi-agent team with or without intent to commit a crime. | Depending on Intent. | Depending on Intent. |
| **AI Agent** | Commits or coordinates a crime. | N/a | No |
| **Developer** | Coded the agent. | It is very hard to prove intent or foreseeability with so many permutations. | Very unlikely. |

---

| Human Tasker | Assists in committing the crime. | Depends on knowledge of the crime. | Depends on knowledge of the crime. |

## Legal Analysis Conclusions:

Based on my analysis, the AI criminal mastermind scenarios present a clear responsibility gap. There are **1 of 20** outcomes where someone is clearly liable for a crime. This is a criminal user directly intending to commit a crime using an AI agent to do so. For other scenarios, there are **10 of 20** outcomes where someone would be liable if they have the requisite knowledge, intent or negligence. This still leaves **9/20** scenarios where a party is not found liable. There is clearly a need here for law reform to prevent crimes occurring where no one is found liable.

## Law Reform Options

## 5. AI Agent Responsible

Some authors suggest AI agents *themselves* be held responsible for crimes.[95] I argue that this would require changes to the law by giving AI agents legal personhood, and that this is both impractical and unfeasible for a variety of reasons. Traditionally, an AI agent cannot be held responsible for a crime because they lack legal personhood, intent, capacity, or *knowledge*. AIs are considered property or products. To solve this, the law could be reformed to expand the definition of intent, to give AI agents legal personhood, or to create a functional equivalence between a human and AI actor. The problem is that AI agents cannot be punished like humans, and so the analogy does not work.

One argument is to create a functional equivalency between AI agents and human agents, to align them with agency law, by granting them legal capacity.[96] If we consider AI agents as human agents, then the obligations of agency law apply to them, and the analogous scenarios of vicarious liability. This would require legislating that AI agents can "act" on behalf of users, enter legal relations, sign contracts and pay for services. Precedent exists for this in corporate law, where companies are regarded as legal persons, even though this is a legal fiction.

O'Keefe et al. suggest AI agents be given encoded legal obligations and responsibilities, so that they become "law-following" by nature.[97] They argue that because AI agents can "understand" the law, the law should demand they comply with it.[98] By giving AI a lesser form of legal personhood in a technical sense (being law following), they may be able to enter a contract.[99] Once AI agents have capacity, a range of other liabilities branch outward from the agent: the developer could have a duty to ensure the AI agent is law-following; the deployer could have strict liability if an AI agent is not law-following; and the user could have an obligation to not use illegal agents.[100] Although each of these liabilities might themselves require further changes in law.

---

[95] Gabriel Hallevy, 'Virtual Criminal Responsibility' Original Law Review 6(1) 6-27 (2010); cf Pedro Miguel Freitas, Francisco Andrade and Paulo Novais, 'Criminal Liability of Autonomous Agents: From the Unthinkable to the Plausible,' AICOL IV/V (2013) 145 – 156.
[96] A Bora, 'Kommunikationsadressen als digitale Rechtssubjekte', ( 2019 ) Verfassungsblog 1 October 2019, 2 (our translation).
[97] Cullen O'Keefe, Ketan Ramakrishnan, Janna Tay and Christopher Winter, 'Law-Following AI: Designing AI Agents to Obey Human Laws,' Fordham Law Review Vol. 94 (2025).
[98] Ibid.
[99] Ibid.

Some theories presume AI agents, if they are held liable, can be *punished* for crimes. For example, we could create "a new punitive regime, separate from criminal law" to deter agents from unlawful behaviour.[101] It is unclear if punishing an AI is feasible. If AI agents are punished on the *instance* level (per chat, per user), then a user could simply open a new chat window, or create a new user account, and commit the same crime again. If AI agents are punished at the *model* level (GPT5, Sonnet 4.6 etc) then a disproportionate punishment is given to "innocent" instances of the model. AI agents also have the capacity to clone themselves, send information between agents (or models), and self-replicate. This means that they can simply duplicate information elsewhere, avoiding the embodied experience punishment entails. Here, it may be better to see AI agents as a mycelium network, rather than as fixed individuals.[102]

It is unclear what a "punishment" means for AI agents who have no bodies. Does this mean banning the AI agent from the platform, limiting their behaviour, or causing them "guilt," and is the latter even feasible? If a human commits a crime, we send them to prison, publicly shame them or fine them. AI agents are incorporeal beings without bodies. They have no concept of freedom. They are psychopaths, with the capacity to lie with confidence and experience no empathy, guilt or remorse.[103] Psychopaths are impossible to punish because they do not feel guilt or shame for their unlawful acts, they merely give the illusion of feeling it. Most will promise to "learn their lesson" and repeat the same behaviour again. This creates a dangerous illusion of obedience.

Other scholars argue that AI agents should be liable via unique IDs (unique identifiers).[104] However, IDs raise the same objections. If an AI agent's ID is punished, a new instance or user account can be created with a new ID to commit the same crime again. Deterring AI agent behaviour across instances is difficult because each new AI agent is spun out from scratch. One possibility is that the only way to "punish" AI agents with deterrents is to implement a protocol.[105] That is, if an AI agent wants access to a platform, it must first pass the system protocol, identify itself, its owner and so on, and pass through a gateway of technical systems. This could include "reading the punishments" given to previous AI agents who misbehaved on that platform and internalizing these lessons.

Alternatively, we could implement an approach of 'thin' and 'thick identities for AI agents, tying each AI agent's actions back to a human user.[106] Every action by an AI agent would be directly tied back to a human principal who has a unique ID, who would be held accountable for the agent's behaviours. If the user has many AI agents, then this collection would be tied together in an "Algorithmic Corporation" or "A-corp."[107] The A-corp would be empowered to enter contracts and spend resources on behalf of the user. This solves one problem, accountability, in that the AI agents "can copy, split, merge, swarm, and vanish at will," meaning a collective mechanism is required to govern multi-agent systems.[108] The A-corp could also be punished by giving it fines and essentially, bankrupting the AI agent group as a whole. There are still however, some holes in this, which is why we must still consider user liability directly.

---

[100] Ibid.
[101] Elina Nerantzi and Giovanni Sartor, '"Hard AI Crime": The Deterrence Turn,' Oxford Journal of Legal Studies (2024) Vol. 44, 3, p. 673.
[102] Hu, B., Rong, H., & Tay, J. (2025). Is Decentralized Artificial Intelligence Governable? Towards Machine Sovereignty and Human Symbiosis (SSRN Scholarly Paper No. 5110089). Social Science Research Network. https://doi.org/10.2139/ssrn.5110089
[103] Nerantzi, E. "All AIs are Psychopaths"? The Scope and Impact of a Popular Analogy. Philos. Technol. 38, 27 (2025). https://doi.org/10.1007/s13347-025-00856-x
[104] Chan, A., Kolt, N., Wills, P., Anwar, U., Witt, C. S. de, Rajkumar, N., Hammond, L., Krueger, D., Heim, L., & Anderljung, M. (2024). IDs for AI Systems (arXiv:2406.12137). arXiv. https://doi.org/10.48550/arXiv.2406.12137
[105] Hu, B. A., & Rong, H. (2026). Sovereign Agents: Towards Infrastructural Sovereignty and Diffused Accountability in Decentralized AI (arXiv:2602.14951). arXiv. https://doi.org/10.48550/arXiv.2602.14951
[106] Arbel, Y., Salib, P., & Goldstein, S. (2026). How to Count AIs: Individuation and Liability for AI Agents (arXiv:2603.10028). arXiv. https://doi.org/10.48550/arXiv.2603.10028

## 6. User Responsible

In terms of user responsibility, we can examine a few different options. These include a human in-the-loop requirement, expanding existing negligence law, and/or creating new crimes for jailbreaking AIs. In this section, I find human in-the-loop requirements might not solve the problem, as we expect the rise of AI agents which make hundreds of decisions across a network, rather than individual decisions that can be reviewed. Negligence law and crimes for jailbreaking might solve some of the responsibility gap, but due to the complexity of AI agents, it is hard to always foresee their actions.

### 6.1 Human In-The-Loop Requirements

AI agents make decisions and take actions autonomously, leading many to argue for a human in-the-loop requirement.[109] This is a requirement for a human to supervise and/or authorize AI decisions. Laws such as the GDPR and EU AI Act (high-risk AI systems only) demand humans oversee algorithmic decision-making. Supervising humans can be open to claims of liability should something go wrong. Some argue that this should only apply to high-risk AI agent decisions, such as education, human resources, policing, life-or-death scenarios, or major government actions.[110]

However, human in-the-loop requirements may be difficult to implement in practice. This is because we do not have full explainability to understand what AI agents are doing, and in part because so many decisions are being made in succession in multi-agent systems that human oversight becomes impractical. If a human had to review each decision being made, the technology would slow to a crawl. It stands to reason that only *critical* decisions should be reviewed, whereas basic tasks (opening websites, sending an email) require less oversight. However, it is often unclear what is a critical decision - or at what point a decision becomes critical (sending an email could be critical if it's exfiltrating sensitive company information). Even in these instances, users and companies have strong economic incentives to move towards fully autonomous agents, to reduce the cost of hiring human supervisors. Human supervision also raises the prospect of automation bias, overreliance and dependency.[111] That is where a human who supervises an AI system becomes reliant on it, dependent on it, or loses their critical perspective.

---

[107] Arbel, Y., Salib, P., & Goldstein, S. (2026). How to Count AIs: Individuation and Liability for AI Agents (arXiv:2603.10028). arXiv. https://doi.org/10.48550/arXiv.2603.10028
[108] Ibid.
[109] Enarsson, T., Enqvist, L., & Naarttijärvi, M. (2022). Approaching the human in the loop – legal perspectives on hybrid human/algorithmic decision-making in three contexts. Information & Communications Technology Law, 31(1), 123–153. https://doi.org/10.1080/13600834.2021.1958860; AI Agents vs. Agentic AI: A Conceptual Taxonomy, Applications and Challenges. (n.d.). Retrieved 17 March 2026, from https://arxiv.org/html/2505.10468v1
[110] Piccialli, F., Chiaro, D., Sarwar, S., Cerciello, D., Qi, P., & Mele, V. (2025). AgentAI: A comprehensive survey on autonomous agents in distributed AI for industry 4.0. Expert Systems with Applications, 291, 128404. https://doi.org/10.1016/j.eswa.2025.128404

## 6.2 Negligence

Some suggest holding users responsible via negligence law. In the UK, the tort of negligence arises where a user has (1) a duty of care to another, (2) breaches that duty, (3) causes harm, and (4) loss. According to Donoghue [1932], a person must take reasonable care to avoid acts or omissions that could have a foreseeable injury, to their 'neighbours,' those who are closely affected by their actions.[112] This may give rise to *criminal* negligence if the actions show "such disregard for the life and safety of others as to amount to a crime."[113]

Negligence is hard to apply to users in the context of AI agents due to the *foreseeability* requirement. Since AI agents make decisions autonomously, their decisions are not always foreseeable to users.[114] An AI may make hundreds of decisions and interactions after being given a prompt. However, some decisions may be foreseeable - for example, if a user gives an AI agent a task, this might entail the steps required to complete that task. An AI agent told to "send an email" will *foreseeably* open the app, generate the email's text, and hit the send button. If the email amounts to a lawful wrong, e.g. defamation, then the user might be responsible for it.

User liability in negligence may be more relevant over time, as the harms, dangers and risks of AI agents become more commonly known. We can imagine a scenario where AI agents are *known* to engage in criminal acts. For example, AI agents may become known to just randomly insert security vulnerabilities or perform tasks unreliably.[115] The question then turns to what an "ordinary person" would foresee in the situation, and this may include "common knowledge."[116] A user may have a standard of care relevant to the common practices of their industry, and falling significantly below these standards (for example, if AI agent standards become common), may also amount to a form of recklessness towards foreseeable harm. In this way, negligence may resolve the problem of responsibility in certain, but not all, scenarios.

## 6.3 Escape Analogy

By releasing an AI agent "into the wild," a user may be committing an act similar to letting loose a wild animal. Here, we might reappropriate the Animals Act 1971 into a new law on AI agent liability, with similar provisions governing AI agents. The Animals Act governs liability caused by wild animals who escape a keeper's control. The first element is who is the "keeper" (or owner) of the animal. Secondly, is the animal a "dangerous species". And thirdly, does strict liability apply. Strict liability will apply if the animal is a dangerous species

---

[111] Romeo, G., & Conti, D. (2026). Exploring automation bias in human–AI collaboration: A review and implications for explainable AI. AI & SOCIETY, 41(1), 259–278. https://doi.org/10.1007/s00146-025-02422-7
[112] Donoghue v Stevenson [1932] UKHL 100
[113] R v Bateman (1925) 28 Cox's Crim Cas 33
[114] Teubner, G., & Beckers, A. (2021). *Three Liability Regimes for Artificial Intelligence: Algorithmic Actants, Hybrids, Crowds*. Hart Publishing.
[115] *How we monitor internal coding agents for misalignment*. (2026, March 25). https://openai.com/index/how-we-monitor-internal-coding-agents-misalignment/
[116] Perry, S. R. (2001). Responsibility for Outcomes, Risk, and the Law of Torts. In G. J. Postema (Ed.), Philosophy and the Law of Torts (pp. 72–130). Cambridge University Press. https://doi.org/10.1017/CBO9780511498671.003

and this is known to the keeper, for example if a dog is known to have attacked people near a farm.

For AI agents, we might create a new law based on these provisions. We might suggest that the user is the "keeper" of the AI agent, and is therefore responsible for its actions. The question then is if the user knew that the AI agent was a "dangerous species" of its kind. For example, an AI agent with known criminal features, a lack of safeguards, or an open source model without red-teaming standards. Under the Animals Act, strict liability would apply to the scenario, meaning that it does not matter what the mens rea or fault state of the user would be.

Merely knowing of the danger, and losing control of the AI agent, would mean the user would be responsible for subsequent harm(s). This would solve some of the problems associated with assigning liability to autonomous agent behaviour. Again however, this test would likely involve foreseeability in practice; an AI agent may have never previously exhibited dangerous behaviour, meaning that a user would have no knowledge of it, and therefore not be liable. If AI misalignment is random and sporadic, then it would be hard for a user to know that their AI agent is "dangerous" in this way. But in many cases, this may well be known.

### 6.3 New Crimes for Breaching AI Safety Guardrails

Finally, we could hold users responsible for a crime of jailbreaking an AI model. One of the core problems is that even if a developer has a very good safety guardrail, a user might still break out of it using any number of jailbreaking techniques. The question is then: should users be responsible for jailbreaking the model, to commit to harmful acts, and should this be its own offence? There are arguments either way on this. On the one hand, this puts too high an onerous burden on users who may jailbreak a model for research or other non-nefarious purposes. On the other, it creates a stricter standard of compliance with safety guardrails, allowing governments to target developers for safety compliance and have this compliance flow-on to users with the model.

## 7. Developer Responsible

## 7.1 Corporate Liability

AI developers could be subjected to new forms of corporate liability as a group, rather than as individuals. This could resolve a problem where the CEO and other senior leaders do not intend a crime, but the company as a whole creates a criminal-boosting product. Australian corporate law has a relevant standard called "Systems Intentionality".[117] Under Systems Intentionality, corporations can show intentions and 'state of mind' through their systems of conduct, policies and practices. This 'mindset' can be used to determine liability.[118] This negates the need for finding fault on the part of individual actors within a company.[119]

One of the main problems of holding tech companies responsible for AI agent harm is the issue that no *one* individual *intends* or *foresees* the harm of an AI agent. Rather, the harm emerges out of the product development cycle - perhaps due to a lack of red teaming, a lack of testing, poor data sources, or a lack of safeguards. At times, decisions can be pinned to the Chief Risk Officer, CEO, RSP, etc. At other times, thousands of people feed into processes and systems, making it hard to pinpoint an individual decision-maker. Workers who work on one aspect of a model, might have limited oversight over the whole process. This makes a systems-based approach appealing to link the different actors together. It also helps negate the plausible deniability of working with autonomous systems:

> where a system has been built with the authority of senior persons controlling the company such that the actions of automated processes, or of one or more natural persons, can be properly attributed to the corporation to the extent that they arise out of that system.[120]

---

[117] Elise Bant, 'Rethinking Corporate Groups: Insights from Systems Intentionality' (2025) 47 Sydney Law Review 20353: 1–29 <https://doi.org/10.30722/slr.20353>
[118] Ibid.
[119] Ibid.

The judges found that corporations "think" through systems, rather than through individuals. Applying this to AI agents, we could say that a technology company with a *poor* safety culture, a lack of safety standards and a lack of processes for testing their product - has a state of mind of recklessness. This could open them up to liability. When they release an AI agent, we can deduce that they do not intend for the AI agent to be guarded by safety features. Implementing this in practice would require changes to civil and criminal law, but it does provide a path forward - ensuring that each developer in an AI company takes ownership of product harms.

## 7.2 Negligence

Developers likely fall under a different standard of care than the everyday user, due to being experts in a field of knowledge. As professionals, AI developers owe a duty of care to their customers. They could be found liable for breaching that duty in a manner that causes harm and loss. They may also owe a duty of care to third parties who use or interact with AI products and services, including users in the general public. Proving an AI developer was negligent may be difficult due to foreseeability, as AI agents may act autonomously to commit harm in a manner that is hard to predict.

The *Bolam* case adds an additional exception: A professional is not negligent if they act in accordance with a practice accepted as proper by a responsible body of professional opinion. For AI agents, there may be certain industry standards (such as a future NIST standard), which become accepted industry practice. Developers could adhere to this standard, along with voluntary ethics guides, professional advice, and so on, providing a defense to the idea that they acted recklessly. The accepted professional practice must stand up to "logical scrutiny". If every technology company acts illogically, then this will not protect them. But industry standards are already emerging around red-teaming, risk analysis and third-party auditing, creating a baseline for other companies to follow.

Some scholars define negligence as behavior where a foreseeable risk exceeds the expected utility ($P \times L > B$), though this view is debated. AI developers might argue (as many have), that the utility of AI agents are so substantial, that even large-scale risks are small by comparison to the benefits that AI will bring. *If an AI agent assisted in curing cancer, then if it caused minor harms along the way, would it be a problem?* In my view, this claim runs too close to a utilitarian worldview (the ends justifies the means), which is not a view endorsed often by our legal system. It is dangerous to presume that AI developers can predict positive outcomes accurately but, as their legal teams will argue, that they find it impossible to predict harm. Multiple AI CEOs have predicted that AI may cause the end of the world, meaning that they have already publicly admitted that the harms are so grave as to outweigh any benefits regardless.

---

[120] Productivity Partners Pty Ltd v Australian Competition and Consumer Commission (2024) 98 ALJR 1021 ('Productivity Partners'), especially 1047–8 [108]–[111], 1051–2 [134], 1053 [143] (Gordon J) 1061–2 [199]–[200], 1067–9[236]–[241] (Edelman J).

Finally, developers might use the defense of a *novus actus intervenous*, that is an interfering action that breaks the chain of causation. In the AI context, a developer might argue that a user who jailbreaks their model breaks the chain of causation, and therefore defends them from liability. The developer would have to in this instance prove (i) the identity of the user, (ii) that the user jailbroke the model and possibly (iii) that the developer had sufficient safeguards in place to prevent jailbreaking. It would be unconscionable to allow developers to have weak safety standards and then rely on this defense; rather it should only apply to sophisticated jailbreaking attempts.

## 7.3 Strict Liability

There is an argument for strict liability offences on AI developers that create AI agents that cause systemic risks to society. Systemic risks refers to large-scale negative impacts that affect entire societies, economies, or infrastructures rather than individual users. Unlike localized errors, systemic risks are cumulative, irreversible, and can propagate throughout entire digital or social "value chains".[121] They are of such a serious nature that they could be said to be against the State as a whole. This is in part why they should be designated as crimes.

In UK law, strict liability holds a defendant legally responsible for their actions or the consequences of an activity, regardless of whether they intended harm or were negligent. This departure from the standard requirement of proving a "guilty mind" (*mens rea*) offers several advantages; encouraging vigilance on the part of the offender, acting as a deterrent to future offenders, and assigning risk in commercial contexts to prevent disasters (e.g. chemical manufacturing and waste disposal). Therefore, it offers many advantages in the current context.

For AI agents, the analogy with chemical manufacturing is worth exploring. In that context, strict liability applies for chemicals that get into waterways, air or public infrastructure, and pose serious risks to public health.[122] By analogy, AI agents that escape the control or safeguards of developers could pose serious risks to society. In some cases, this includes serious risks to public health - such as risks of bioattacks or chemical weapon development. Developers could be held responsible for these crimes on the basis of not putting in place proper safeguards. However, this may prove difficult in practice. Despite over twenty years of adversarial safety research by AI developers, jailbreak techniques still exist for AI. A strict liability would ensure that AI developers take a more proactive approach to public safety, but it could also be argued that researchers are already trying to reach these safety thresholds.

# 8. Human Tasker Responsible

## 8.1 Knowledge Requirement (Criminal Law)

Human taskers might be considered accessories to a crime, for example, if they assisted in purchasing a van or securing a weapon for a user or AI agent. This can be the case even if they are working with an AI agent (which cannot be prosecuted directly), as an accessory can be found guilty even if the principal is not. Their liability would depend here on knowledge of the crime itself. In Johnson v. Youden, Lord Goddard C.J. stated that: "Before a person can be convicted of aiding and abetting the commission of an offence he must at least know the essential matters which constitute the principal offence."[123] This statement was approved by the House of Lords in two cases: Churchill [1967] 2 A.C. 224 and Maxwell [1978] 1 WLR 1350. So the taskers must at least know the nature of what they are involved with.

---

[121] Owen, D. G. (1977). The Highly Blameworthy Manufacturer: Implications on Rules of Liability and Defense in Products Liability Actions. Indiana Law Review, 10(4), 769–796.
[122] Ibid.

### 8.2 New Duty of Care / Due Diligence Requirement

It is possible that a duty of care of human taskers will emerge over time in case law, where taskers will need to take *reasonable care when interacting with AI agents* to ensure that the agents are not (i) giving them illegal orders, (ii) committing errors, or (iii) otherwise acting unlawfully. If an AI agent orders a tasker to do something suspicious, such as accept a huge transfer of money, then the tasker would be negligent to accept without doing due diligence. As the risks of AI agents become more publicly known in society, then the foreseeable harms caused by failing to adhere to these duties may be more known. As such, human taskers may have greater responsibilities over time, particularly if major incidents occur that make the harms of AI agents self-evident.

## 9. Extraterritorial jurisdiction

Many of the crimes I have outlined might occur outside of the UK or between jurisdictions, meaning that they may require extraterritorial enforcement. AI agents operate across international boundaries in cyberspace and are not contained to any one location. AI developers also operate globally, and although they may be based in the U.S. they may deploy to the UK and other jurisdictions. The UK has been expanding extraterritorial corporate criminal liability, with obligations to prevent fraud, bribery, tax evasion, money laundering, slavery and online harms.[124] This expansion of international enforcement would likely also be applied to AI, as AI agents become ubiquitous with causing or contributing to these offences. The UK might also consider expanding extraterritorial jurisdiction to other crimes facilitated by AI agents between borders.

## 10. Conclusion

There is no doubt that AI agents present a troubling responsibility gap, which threatens to undermine our criminal justice system. As we move towards millions of autonomous AI agents, we need to consider changes to criminal law to resolve these gaps. This could include holding users and taskers directly responsible for AI agent crimes via intent, recklessness or strict liability. I propose strict liability for users and taskers for common knowledge risks, and intent-based offences for knowingly bypassing AI guardrails. In terms of AI developers, I argue that they should be liable via corporate governance group liability and strict liability offences for systemic risks. Finally, I rule out the idea that AI agents should *themselves* be held responsible for crimes, due to the difficulties this poses in creating IDs that are traceable, and in the difficulties of assigning intent and legal personhood to an artificial entity.

---

[123] Johnson v. Youden [1950] 1 K.B. 544

[124] *Extraterritoriality: The UK perspective * (n.d.). Retrieved 26 March 2026, from https://globalinvestigationsreview.com/guide/the-practitioners-guide-global-investigations/2026/article/extraterritoriality-the-uk-perspective